\documentclass[
    ,final            % use final for the camera ready runs
  ]
  {aipproc}
\pdfoutput=1
\layoutstyle{6x9}

\begin{document}

\title{Fluctuation relations without micro-reversibility for two-terminal conductors}


\author{Heidi F\"orster}{
  address={Institute for Environment and Human Security, United Nations University, UN-Campus, Hermann-Ehlers-Str.~10, D-53113 Bonn, Germany}
}

\author{Markus B\"uttiker}{
   address={Department of Theoretical Physics, University of Geneva,
  CH-1211 Geneva 4, Switzerland}
}

\begin{abstract}
In linear transport, the fluctuation-dissipation theorem relates 
equilibrium current correlations to the linear conductance coefficient. Theory and experiment have shown that in small electrical conductors the non-linear I-V-characteristic of two-terminal conductor 
exhibits terms which are asymmetric in magnetic field and thus micro-reversibility is manifestly broken. We discuss a non-equilibrium fluctuation dissipation theorem which is not based on micro-reversibility. 
It connects the antisymmetric nonlinear conductance with the third cumulant
of equilibrium current fluctuations and a noise term that is proportional to
temperature, magnetic field and voltage. 
\end{abstract}

\maketitle

%%%%%%%%%%%%%%%%%%%%%%%%%%%%%%%%%%%%%%%%%%%%
%% MAINMATTER
%%%%%%%%%%%%%%%%%%%%%%%%%%%%%%%%%%%%%%%%%%%%

\subsection{Linear and nonlinear transport coefficients}
The linear transport regime is governed by microscopic reversibility. Based on this principle, Onsager derived the symmetry of transport coefficients of irreversible processes \cite{onsager}. For electrical conductance, this means that the linear conductance $G_1$ of a two-terminal conductor is an even function of magnetic field.
Another consequence of micro-reversibility is the fluctuation dissipation theorem \cite{einstein} which states that the equilibrium fluctuations $S_{eq}$ are proportional to temperature and to the linear conductance \cite{nyquist} ($k_B$ is the Boltzmann constant),
\begin{equation}\label{S0}
G_1(B)\,=\,G_1(-B),\,\hspace{3mm} S_{eq}\,=\,2k_BT\,G_1.
\end{equation}
Eqs.~(\ref{S0}) are cornerstones of linear transport theory \cite{kubo}. 

The question whether there exist fluctuation relations which apply beyond the linear transport regime has long been of interest \cite{stratonovich}. Recently the question was raised specifically for mesoscopic conductors in the context of theoretical works that characterizes transport not only by conductance and noise \cite{but92} but in terms of the full counting statistics \cite{levitov}. An early discussion is provided by Tobiska and Nazarov \cite{tobiska} and followed by different discussions \cite{andrieux,espositoPRB,astumian,saito}. In particular Saito and Utsumi \cite{saito} proposed a fluctuation relation in the presence of a magnetic field. Their derivation assumes that micro-reversibility also holds beyond the linear transport regime. However it has been known both from theoretical works \cite{sanchez,spivak,polianski06,glazman} and from experiments \cite{zumbuhl,wei,marlow,leturcq,angers, chepelianskii} that the current proportional to the square of the voltage (the rectification coefficient) is in general not an even function of magnetic field and thus  manifestly breaks micro-reversibility. Therefore, it is interesting to ask whether non-equilibrium fluctuation relations exist even in the case when there is a manifest departure from micro-reversibility. The surprising answer is, yes, even for samples with magnetic field asymmetry, there exist fluctuation relations \cite{forster08}.

We now emphasize the leading order behavior in temperature $k_BT$, magnetic field $B$, and voltage $V$. 
Expanding the current $I$ through a mesoscopic system up to the first nonlinear contribution defines the nonlinear conductance coefficient $G_2=[\partial^2 I/\partial V^2]_{eq.}$, 
\begin{equation}
	I\,=\,G_1\,V\,+\,(1/2)\,G_2\, {V^2}\,+\,\ldots\,,\hspace{3mm}	G_A\,\sim\, B.
\end{equation}
The nonlinear transport coefficient $G_2$ can be decomposed in a part $G_S$ that is an even function of magnetic field and exists also for non-interacting conductors and a part $G_A$ that is asymmetric in magnetic field and that exists only for interacting conductors, $G_{S/A}=G_2(B)\pm G_2(-B)$. For a weak field the antisymmetric part is $G_A \sim\, B$.
In chaotic cavities $G_2$ is a consequence of quantum interference and the coefficient fluctuates from sample to sample. It is zero on average. It's mean squared value for a magnetic flux larger than a flux quantum is given by Sanchez and one of the authors. The entire crossover from low to high magnetic field is discussed by Polianski and one of the authors~\cite{polianski06} and is illustrated in Fig. 1. The crossover flux $\Phi_c$ occurs at magnetic fields generating a flux through the sample much smaller than the flux quantum $\Phi_0 = h/e$. Experiments on chaotic cavities were performed by Zumb\"uhl et al.~\cite{zumbuhl}. Experiments on 
rings~\cite{leturcq,angers} and carbon nanotubes~\cite{wei} demonstrate that magnetic field asymmetry is generic. It has been investigated also for multi-terminal Hall bars \cite{chepelianskii}. While the work on chaotic cavities emphasizes the quantum nature of the effect, at high magnetic fields orbital effects can lead to a classical magnetic field asymmetry~\cite{glazman}.

\begin{figure}
\includegraphics[height=5.5cm]{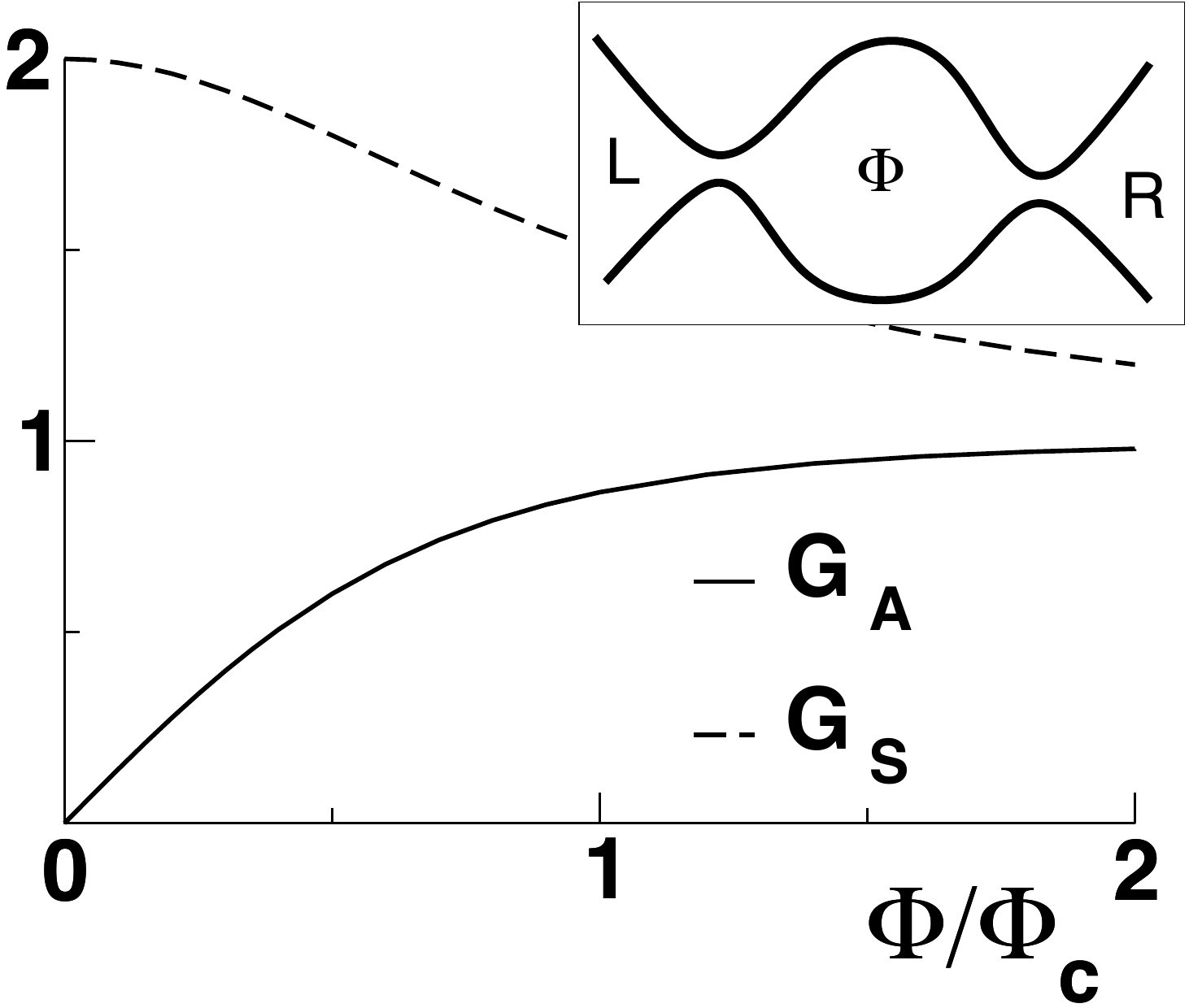}\hspace{5mm}
\includegraphics[height=5cm]{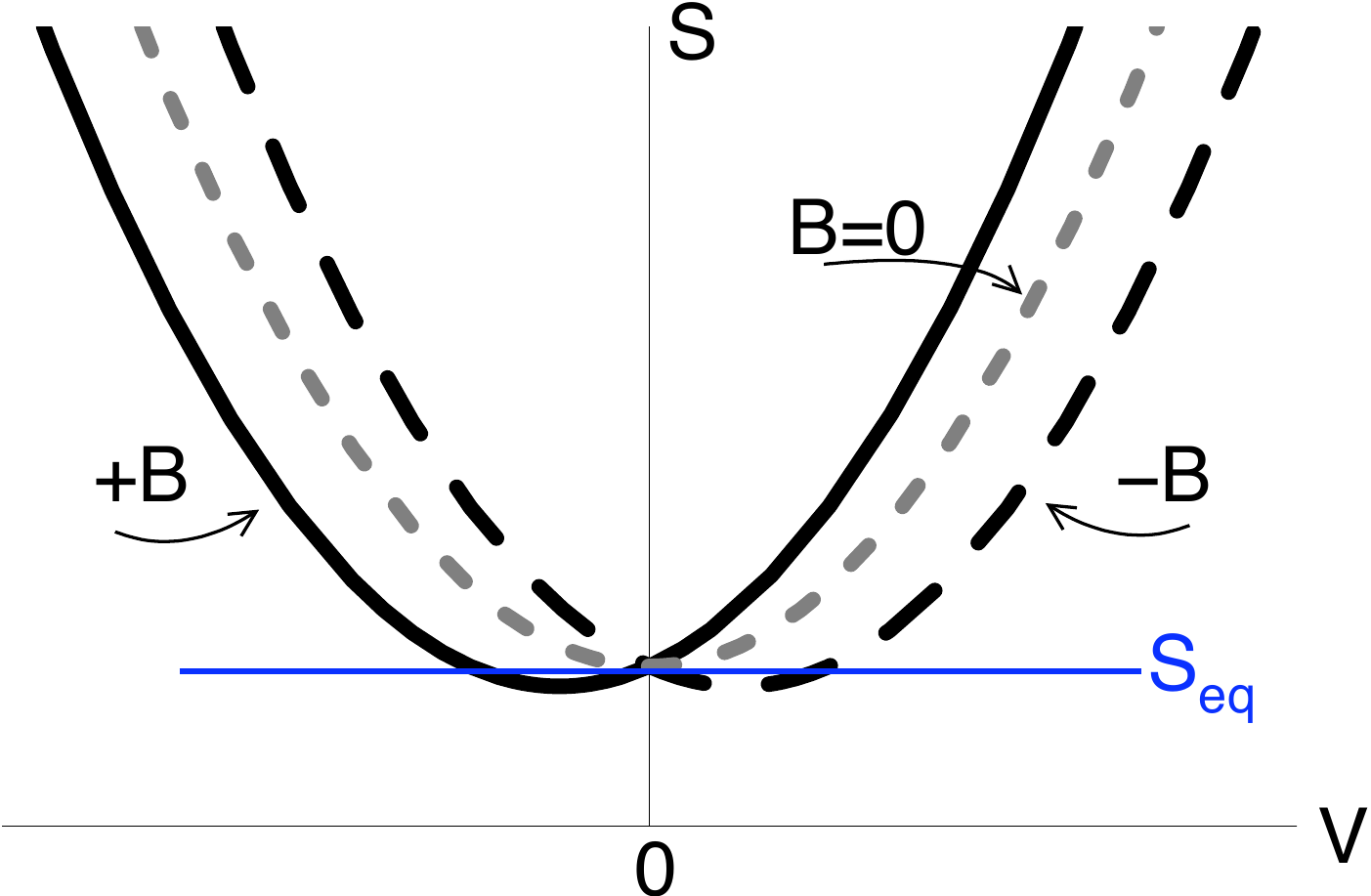}
\caption{Left: The root mean squared symmetric $G_S$ and antisymmetric $G_A$ rectification coefficients of a chaotic quantum dot with two contacts (inset) as a function of flux through the dot. $\Phi_c = h/e$ is the crossover flux for the transition from 'low' to 'high' magnetic field (after Ref. \protect\cite{polianski06}). Right: Sketch of the noise as a function of voltage; for finite magnetic field it can drop below the Nyquist-Johnson noise $S_{eq}$.
}
\label{exp}
\end{figure}

The magnetic field asymmetry of the nonlinear coefficient implies the {\it absence of microscopic reversibility out of equilibrium}. Large external voltages modify the electron density within the conductor, which is subject to Coulomb interaction. In other words, the local internal potential $U=U(\vec r,V)$ of the conductor responds to the shifts of the applied voltage $V$  and has to be determined self-consistently. Crucially, the internal potential is not necessarily an even function of the magnetic field $B$, but has both symmetric and anti-symmetric components. Therefore, scattering from left to right at $+B$ does not occur with the same probability as the corresponding process from right to left at $-B$: This is nothing else but a lack of micro-reversibility out of equilibrium.

\subsection{Fluctuation relation for nonlinear transport}
We could expect that with the absence of micro-reversibility, fluctuation relations similar to (\ref{S0}) would also not exist. However, this is not the case. Just like the current, also the noise $S$ can be expanded around equilibrium for $eV \ll k_BT$, defining the linear and quadratic coefficients $S_1$ and $S_2$,
\begin{equation}
	S\,=\,S_{eq}\, + \,S_1 \,V\,+(1/2)\,\,S_2 \,{V^2} \,+\,\ldots\,\,\,\,
\end{equation}
The noise susceptibility $S_1=[\partial S/\partial V]_{eq.}$ is the deviation from equilibrium noise {\it linear} in voltage. This contribution contains emerging shot noise. As for the nonlinear conductance, we define the (anti-)symmetric noise susceptibility $S_{S/A}=S_1(B)\pm S_1(-B)$, as well as the (anti-)symmetric third cumulant at equilibrium $C_{S/A}^{eq.}=C_3^{eq.}(B)\pm C_3^{eq.}(-B)$.  Surprisingly, one can derive nonlinear fluctuation relations without making use of microscopic reversibility \cite{forster08}. 
These relations imply on the one hand that the symmetric third cumulant vanishes at equilibrium which means that it is odd in magnetic field, and that   the symmetric linear coefficient of noise is proportional to temperature and to the nonlinear conductance coefficient. On the other hand it states that the antisymmetric third cumulant is composed of the corresponding linear noise coefficient and nonlinear conductance, 
\begin{equation}\label{S1}
	S_S\, =\, k_BT\,G_S, \hspace{3mm} C_S^{eq.}\,=\,0, \hspace{3mm} C_A^{eq.}\,=\,3k_BT(S_A-k_BTG_A). 
\end{equation}
The third cumulant at equilibrium vanishes if the transmission through the conductor does not depend on energy. Taking into account the first order correction linear in energy, the third cumulant is proportional to $(k_BT)^2$.
The magnetic field symmetry of the third cumulant at equilibrium is a consequence of micro-reversibility at equilibrium: One can show \cite{forsterarX}, that this implies even (odd) cumulants at equilibrium to be even (odd) in magnetic field, in particular $C_A^{eq.}\sim B$ for weak fields. 
Altogether, this has the following  consequences: First, the symmetric noise susceptibility $S_S$ has the same $B$-field dependence as the nonlinear conductance $G_S$. Second, the antisymmetric contribution to noise is proportional to magnetic field, temperature and voltage  for weak fields \cite{forster08,sanchez09},
		\begin{equation}
			 S_A\,\sim\, k_B T\, B\, V.
		\end{equation}
Thus in the presence of electron-electron interaction the noise is not an even function of magnetic field and voltage (see Fig.~\ref{exp}). 
Note that there exists a range of voltages for which the non-equilibrium noise is below the Johnson-Nyquist noise \cite{lesovik}.

We emphasize that fluctuation relations have been derived before for systems without magnetic field \cite{tobiska, andrieux} and also in the presence of a magnetic field \cite{saito}.  The symmetric fluctuation relations in Eqs.~(\ref{S1}) are identical to those from Ref.~\cite{saito} which means that they remain valid even in the absence of microscopic reversibility. The antisymmetric relations however differ:  A derivation~\cite{saito} based on micro-reversibility leads to the conclusion that $G_A$ and $S_A$ in a two-terminal system both are proportional to the asymmetric third cumulant,  $C_A^{eq.}\sim G_A\sim S_A$, which means that they are  even functions of magnetic field for systems with a vanishing third cumulant. In contrast our derivation of the fluctuation relation in the absence of micro-reversibility  permits terms asymmetric in magnetic field for the nonlinear conductance and the noise susceptibility even if $C_3^{eq.}=0$.

\subsection{Derivation}
The importance of the fluctuation relation and also its generality require a careful discussion of the derivation. 
It is possible to derive the fluctuation relations from the full counting statistics without specifying any model of interaction.
The full counting statistics of a two-terminal conductor is the probability distribution 
$P(Q)$ that a number of $Q$ charges are transmitted into the  reservoirs during the measurement time
$t$. The distribution function $P(Q)$ is expressed by the generating function $F(i\lambda)=\ln\sum_Q P(Q)e^{i\lambda Q}$,
where the counting field $\lambda$ is the conjugate variable to $Q$. In the long time limit, all zero-frequency cumulants of the current can be expressed using derivatives of the generating function with respect to the counting field, evaluated at $\lambda=0$. The mean current and the noise are given by derivatives of $F$ with respect to the counting field, $I=(e/it)[\partial F/\partial\lambda]_{\lambda=0}$,  $S=(e^2/i^2t)[\partial^2 F/\partial\lambda^2]_{\lambda=0}$, thus  the rectification coefficient, the noise susceptibility and the third cumulant at equilibrium are explicitly
\begin{equation}
	G_2=\left.\frac{e}{it}\frac{\partial^3 F}{\partial\lambda\partial V^2}\right|_{0}, \hspace{3mm}
	S_1=\left.\frac{e^2}{i^2t}\frac{\partial^3 F}{\partial\lambda^2\partial V}\right|_{0}, \hspace{3mm}
	C_3^{eq.}= \left.\frac{e^3}{i^3t}\frac{\partial^3 F}{\partial\lambda^3}\right|_{0},
	\label{coeff}
\end{equation}
with the index $0$ meaning $\lambda=V=0$. 
We assume the temperature in the left and right reservoir to be equal, and call the ratio $A=eV/k_BT$ the affinity of the system, with $V=V_L-V_R$ the applied voltage. A magnetic field $B$ perpendicular to the conductor is externally controlled.

It is convenient to use the notation $F=F(i\lambda, A)$ in order to emphasize the dependence of the generating function on the affinity. A derivation of the nonlinear fluctuation relation is based on the following properties of the generating function:
\begin{eqnarray}
	F(0,A)\,=\,0, \hspace{3mm}
	F(-A,A)\,=\,0, \hspace{3mm} F(i\lambda,0,+B)=F(-i\lambda,0,-B).
\label{FA}
\end{eqnarray}
The first equation represents conservation of particles, which in terms of the distribution function is expressed by $\sum_QP(Q)=1$. More subtle is the second identity. It defines a special symmetry point of the generating function, which translates for the distribution function into a global detailed balance relation $\sum_Q P(Q)e^{AQ}=\langle{e^{AQ}}\rangle=1$. Importantly, this can be derived without making use of micro-reversibility, and for arbitrary electron-electron interaction. The only assumptions made are that the system consisting of conductor and leads is at all times described by the number of particles in the different components, and that both the total energy and the total number of particles is conserved. For a derivation see Ref.~\cite{forsterarX}. Only the third equation in (\ref{FA}) makes a statement on the  magnetic field symmetry of the generating function,  it corresponds to the micro-reversibility at equilibrium. From this follows, that even (odd) cumulants at equilibrium are even (odd) in magnetic field.

The fluctuation relation (\ref{S1}) can be obtained from the above identities. To this end, consider the generating function with variables $i\lambda$ and $-i\lambda-A$, expanded into Taylor series around $i\lambda=A=0$:
\begin{eqnarray}
	F(i\lambda,A)&=&\sum_{kl}f_{kl}	\frac{(i\lambda)^kA_\alpha^l}{k!l!} \label{ft_taylor1}\\   
	F(-i\lambda-A,A)&=&\sum_{kl}f_{kl}	\frac{(-A-i\lambda)^kA^l}{k!l!}=
	\sum_{pq}\tilde f_{pq} 	\frac{(i\lambda)^pA^q}{p!q!} \label{ft_taylor2}   
\end{eqnarray}
The Taylor coefficients are given by
$f_{kl} =[\partial^{k+l}F( i\lambda,A)/d(i\lambda)^kdA^l]_0$ and 
$\tilde f_{pq} =[\partial^{p+q}F(-i\lambda-A,A)/d(i\lambda)^pdA^q]_0$. 
As indicated by Eq.~(\ref{coeff}), the coefficients $f_{kl}$ are  directly proportional to response coefficients.
From Eq.~(\ref{ft_taylor2}),  a relation between the
coefficients  $\tilde f_{pq}$ and $f_{kl}$ is defined: 
\begin{equation}\label{ft_sstilde}
	\tilde f_{pq}=\sum_{n=0}^q{q\choose n} (-1)^{p+n} f_{p+n,q-n}
\end{equation}
Now the symmetries defined by Eqs.~(\ref{FA}) are crucial, they determine that the coefficients $f_{0q}$ and  $\tilde 
f_{0q}$ all vanish. Setting $p=0$ in Eq.~(\ref{ft_sstilde}) and separating the first and the last term in the sum one obtains
\begin{equation}\label{ft_relationf}
	f_{q0}=-\sum_{n=1}^{q-1}{q\choose n} (-1)^{n} f_{n,q-n}.
\end{equation}
This actually represents a multitude of fluctuation relations, relating different response coefficients of current cumulants of different order. More details on this, and a graphical representation of the fluctuation relations are given in Ref.~\cite{forster08}. Here, we concentrate on the first nonlinear fluctuation relation, Eq.~(\ref{S1}), which follows from the above identity by setting $q=3$: 
\begin{equation}
	f_{30}=3(f_{21}-f_{12}).
	\end{equation}
Only from the third equation in Eqs.~(\ref{FA}), the magnetic field symmetry of the  Taylor coefficients $f_{q0}$ is determined. It says in particular that the symmetric part $f^S_{30}$ vanishes, but makes no statement on the asymmetric contribution. 	Identifying the Taylor coefficients $f_{12}$ and $f_{21}$ as proportional to the nonlinear conductance and the noise susceptibility respectively and (anti-)symmetrizing, the fluctuation relations Eqs.~(\ref{S1}) are directly obtained.
It is important to notice that the nonlinear fluctuation relations take on the form (\ref{S1}) only for a two-terminal conductor. The case of a conductor with multiple terminals is discussed in Ref.~\cite{forster08}.

\subsection{Conclusion} 
The linear transport regime is governed by Onsager-Casimir relations and the fluctuation-dissipation theorem, both derived from the principle of microscopic reversibility at equilibrium. In the nonlinear transport regime, interactions can not be neglected anymore. They can lead to a lack of micro-reversibility out of equilibrium and in the presence of a magnetic field. A consequence is a contribution to the mean current proportional to magnetic field and quadratic in voltage as well as a contribution to the current noise linear in magnetic field and in voltage. 
We have shown that even in this case fluctuation relations exist:
interestingly a noise contribution proportional to temperature, magnetic field and
voltage is linked to the asymmetric nonlinear conductance and the third
cumulant of equilibrium current fluctuations. 
To our knowledge, this relation has not yet been experimentally verified. Mesoscopic physics with its highly controlled and accurate experiments would seem especially suited to demonstrate a non-equilibrium fluctuation relation. 

\subsection{Acknowledgments} 

This work is supported by the Swiss NSF, MaNEP and the European STREP project SUBTLE. 
We thank D. Astumian and Y. Utsumi for correspondence and discussions.


\begin{thebibliography}{99} 


\bibitem{onsager}
L. Onsager, Phys. Rev. {\bf 37}, 405 (1931); H. B. G. Casimir, Rev. Mod. Phys. {\bf 17}, 343 (1945).

\bibitem{einstein}
A. Einstein, Ann. Phys. Lpz. {\bf 17}, 549 (1905).

\bibitem{nyquist}
H. Nyquist, Phys. Rev. {\bf 32}, 110 (1928); J. B. Johnson, Phys. Rev. {\bf 32}, 97 (1928).

\bibitem{kubo}
R. Kubo, J. Phys. Soc. Japan {\bf 12}, 570 (1957).

\bibitem{stratonovich}
G. N. Bochkov and Yu. E. Kuzovlev, Physica A {\bf 106}, 443 (1981). 

\bibitem{but92}
M. B\"{u}ttiker, Phys. Rev. B {\bf 46}, 12485 (1992).

\bibitem{levitov}
L. S. Levitov and G. Lesovik, JETP Lett. {\bf 58}, 230 (1993).

\bibitem{tobiska}
J. Tobiska and Yu. V. Nazarov, Phys. Rev. B {\bf 72}, 235328 (2005); see
especially Sec.~V.

\bibitem{andrieux} 
D. Andrieux, and P. Gaspard, J. Stat. Mech. P01011 (2006);
J. Stat. Mech. P02006 (2007); D. Andrieux, P. Gaspard, T. Monnai, S. Tasaki, arXiv:0811.3687  

\bibitem{espositoPRB}
M. Esposito, U. Harbola, and S. Mukamel, Phys. Rev. B {\bf 75}, 155316 (2007).

\bibitem{astumian}
R. D. Astumian, Phys. Rev. Lett. {\bf 101}, 046802 (2008);
Phys. Rev. E {\bf 79}, 021119 (2009).

\bibitem{saito}
K. Saito and Y. Utsumi, arXiv:0709.4128 and Phys. Rev. B {\bf 78}, 115429 (2008);
Y. Utsumi and K. Saito, arXiv:0810.1113 


\bibitem{sanchez}
D. Sanchez and M. B\"uttiker, Phys. Rev. Lett. {\bf 93}, 106802 (2004).

\bibitem{spivak}
B. Spivak and A. Zyuzin, Phys. Rev. Lett. {\bf 93}, 226801 (2004).

\bibitem{polianski06}
M. L. Polianski and M. B\"{u}ttiker, Phys. Rev. Lett. {\bf 96}, 156804 (2006).

\bibitem{glazman}
A. V. Andreev and L. I. Glazman,  Phys. Rev. Lett. {\bf 97}, 266806 (2006).

\bibitem{zumbuhl}
D. M. Zumb\"uhl et al., Phys. Rev. Lett. {\bf 96}, 206802 (2006).

\bibitem{wei}
J. Wei et al., Phys. Rev. Lett. {\bf 95}, 256601 (2005).

\bibitem{marlow}
C. A. Marlow et al., Phys. Rev. Lett. {\bf 96}, 116801 (2006). 

\bibitem{leturcq}
R. Leturcq et al., Phys. Rev. Lett. {\bf 96}, 126801 (2006).

\bibitem{angers}
L. Angers et al., Phys. Rev. B {\bf 75}, 115309 (2007).

\bibitem{chepelianskii}
A. D. Chepelianskii and H. Bouchiat, Phys. Rev. Lett. {\bf 102}, 086810 (2009).

\bibitem{forster08}
H. F\"orster and M. B\"uttiker, Phys. Rev. Lett. {\bf 101}, 136805  (2008).

\bibitem{sanchez09} 
D. Sánchez, Phys. Rev. B {\bf 79}, 045305 (2009).

\bibitem{lesovik}
G. B. Lesovik and R. Loosen, Z. Phys. B {\bf 91},  531 (1993). 

\bibitem{christen} 
T. Christen and M. B\"uttiker, Europhys. Lett. {\bf 35} (7), 523 (1996).

\bibitem{forsterarX}
See appendix of H. F\"{o}rster and M. B\"{u}ttiker, cond-mat:0805.0362 




\end{thebibliography}
\end{document}